\documentclass[reprint,amsmath,amssymb,aps,preprintnumbers,nofootinbib,showpacs]{revtex4-1}

\usepackage{amssymb}
\usepackage{graphicx}
\usepackage{subfigure}
\usepackage{fullpage}
\usepackage{hyperref}

\newcommand{\mone}{$M_1$}
\newcommand{\mtwo}{$M_2$}

\newcommand{\missET}{E_{\rm T}^{\rm miss}}
\usepackage{tikzfeynman}   % Flip's Feynman Diagrams

\begin{document}

\preprint{UCI-TR-2015-15}

\title{Limiting SUSY compressed spectra scenarios}

\begin{abstract}
 Typical searches for supersymmetry cannot test models in which the two lightest particles have a small (``compressed'') mass splitting, due to the small momentum of the particles produced in the decay of the second-to-lightest particle. However, datasets with large missing transverse momentum ($\missET$) can generically search for invisible particle production and therefore provide constraints on such models. We apply data from the ATLAS mono-jet (jet+$\missET$) and vector-boson-fusion (forward jets and $\missET$) searches to such models. The two datasets have complementary sensitivity, but in all cases experimental limits are at least five times weaker than theoretical predictions.
\end{abstract}

\author{Andy Nelson}
\affiliation{Department of Physics and Astronomy, UC Irvine, Irvine, CA 92627}
\author{Philip Tanedo}
\affiliation{Department of Physics and Astronomy, UC Irvine, Irvine, CA 92627}
\author{Daniel Whiteson}
\affiliation{Department of Physics and Astronomy, UC Irvine, Irvine, CA 92627}

\date{\today}

\maketitle
 
\section{Introduction}

Models of supersymmetry (SUSY) remain leading candidates for new physics despite increasingly stringent constraints from collider searches. In particular, the lightest SUSY particle (LSP, also denoted as $\chi^0_1$) is a prototypical candidate for `weakly interacting massive particle' (WIMP) dark matter. 
Conventional searches for SUSY require that the mass splitting between the invisible LSP and the next-to-lightest SUSY particle is large enough to produce a visible probe, such as a charged lepton.   However, many interesting SUSY models have a splitting which is very small, known as {\emph {compressed spectra}}, to which these limits do not apply.

%Models of supersymmetry (SUSY) remain leading candidates to solve the hierarchy problem and provide a dark matter candidate, despite increasingly stringent constraints from collider searches.  However, most searches for SUSY require that the mass splitting between the invisible lightest SUSY particle (LSP, also denoted as $\chi^0_1$) and the next-to-lightest SUSY particle is large enough to produce a visible probe, such as a charged lepton.   However, many interesting SUSY models have a splitting which is very small, known as {\emph {compressed spectra}}, to which these limits do not apply.

% Theory stuff

In the minimal supersymmetric Standard Model (MSSM)\footnote{See, e.g.~\cite{Csaki:1996ks}~(\cite{Rosiek:1989rs}) for a brief (comprehensive) review.}, the neutralino is an admixture of the neutral Higgsinos and gauginos. 
To obtain a realistic dark matter candidate, one must typically balance the large annihilation rates of the Higgsinos ($\tilde h_{u,d}$) or SU(2) gauginos (winos) against the small annihilation rates of the hypercharge gaugino (the bino, $\tilde B$) to yield the correct dark matter relic abundance through thermal freeze-out; this tuning of admixtures is called a `well-tempered neutralino'~\cite{welltempered}. 
This scenario is naturally associated with a compressed spectrum. In the case of a bino--Higgsino LSP, the tuning of the $\tilde B$ and $\tilde h_{u,d}$ masses typically gives a splitting between the LSP and the next-lightest states of $\mathcal O(M_Z)$.

%In such models, the dark matter candidate $\chi^0_1$ is an admixture of the hypercharge gaugino (the bino, $\tilde B$) and the neutral Higgsinos ($\tilde h_{u,d}$).  This is an example of a `well-tempered neutralino' wherein the large annihilation rates of the Higgsino balances the small annihilation rates of the bino to yield the correct dark matter relic abundance through thermal freeze-out~\cite{welltempered}. More generally, this scenario is naturally associated with a compressed spectrum\footnote{In the case of a well-tempered bino--wino LSP, this compression allows for co-annihilations in the early universe.}: the tuning of the bino and Higgsino masses typically gives a splitting between the LSP and the next-lightest states of $\mathcal O(M_Z)$.

% What this paper is

In this paper, we probe this experimentally challenging yet theoretically important region using two broadly powerful datasets from the ATLAS experiment: the ``mono-jet''\cite{atlasMonojet} final state (one or more jets with large missing transverse momentum) and the dataset used to search for vector-boson-fusion (VBF) production of a Higgs boson~\cite{atlasVBF} which decays invisibly (two forward jets and large missing transverse momentum). 
Previously, analyses have applied the mono-jet limits to compressed spectra scenarios with electroweakinos~\cite{Schwaller:2013baa, Han:2013usa, Han:2014xoa}, phenomenological MSSM~\cite{Aad:2015baa}, or simplified models of the MSSM~\cite{Arbey:2015hca}; or calculated projected LHC sensitivity to VBF scenarios~\cite{giudice, Dutta:2012xe, Delannoy:2013ata}. To our knowlegde, this is the first application of the VBF dataset to these SUSY scenarios. 

These data are applied to evaluate the constraints on supersymmetric models with compressed spectra in which all supersymmetric partners except the electroweak gauginos are very massive~\cite{welltempered,giudice,sinha}.
We do not impose that the thermal relic abundance of the LSP matches that of dark matter and allow the possibility of non-thermal production or multi-component dark matter \cite{Gelmini:2010zh,Cheung:2012qy}. Direct detection bounds are discussed in \cite{Cheung:2012qy}, where it is also noted that the well-tempering region considered in this work below overlaps with a blind spot in spin-independent direct detection constraints.

\section{Theory} 

We review the main features of bino--Higgsino dark matter in the well-tempered, blind-spot limit following the discussions in \cite{welltempered} and \cite{Cheung:2012qy}. This is a region of the Minimal Supersymmetric Standard Model where we take the limit where the gluino and the scalar superpartners of Standard Model fermions are decoupled so that (1) they play no role in the dynamics of the dark matter and (2) the models are not excluded by colored sparticle searches from Run I. We shall further take the case where the SU(2) gaugino mass and CP-odd Higgs mass are taken very large compared to the other electroweak fermionic sparticle mass parameters, $M_2,M_A \gg M_1, |\mu|$ so that the winos and additional Higgses decouple. 
We are left with a theory that extends the Standard Model with two charged fermions $\chi_1^\pm$ of mass  $M_{\chi_1^\pm}\approx |\mu|$ and three neutral fermion states, $\tilde B$, $\tilde h_{u,d}$ which mix into three mass eigenstates $\chi^0_{1,2,3}$, the lightest of which is identified with dark matter~\cite{Feng:2000gh, Feng:2000zu, Baer:2002fv, Giudice:2004tc, Pierce:2004mk, Masiero:2004ft, welltempered, Baer:2006te}.  The parameters of this theory are assumed real and are the bino mass, $M_1>0$, the Higgsino mass, $\mu$ (taking either sign), and the ratio of the Higgs vacuum expectation values, $\tan \beta$.  We further assume $M_1, |\mu| > M_Z$. 

\subsection{Well Tempering and Compressing}

The well-tempering condition in the bino--Higgsino theory described above is that $M_1 \approx |\mu|$. 
It is useful to go to an intermediate basis of states,
\begin{align}
	\tilde B 
	&&  \tilde h_- \equiv \frac{\tilde h_u - \tilde h_d}{\sqrt{2}}
	&&  \tilde h_+ \equiv \frac{\tilde h_u + \tilde h_d}{\sqrt{2}}.
	\label{eq:well:temper:b:hm:hp}
\end{align}
In this basis, the neutralino mass matrix is
\begin{align}
%	M_{\chi^0} = 
	\begin{pmatrix}
		M_1 
		& -\frac{s_\beta+c_\beta}{\sqrt{2}}s_{W} M_Z 
		& \frac{c_\beta-s_\beta}{\sqrt{2}}s_W M_Z 
		\\
		-\frac{s_\beta+c_\beta}{\sqrt{2}}s_W M_Z 
		& \mu
		& 0
		\\
		\frac{s_\beta-c_\beta}{\sqrt{2}}s_W M_Z 
		& 0
		& -\mu
	\end{pmatrix}
\end{align}
up to higher order terms suppressed by $M_W^2/M_2 \ll M_Z$ and where $s_\beta$ and $c_\beta$ are $\sin \beta$ and $\cos \beta$ respectively. %Observe that $\tan\beta$ controls the relative mixing with the different Higgsinos. 

We are interested in the regime where the spectrum is compressed,  
\begin{align}
\left|M_1\pm \mu\right| < \frac{s_\beta\mp c_\beta}{\sqrt{2}}s_W M_Z
\label{eq:well:tempered:compression}
\end{align}
for $M_1 \approx \mp \mu$. In this case, a pair of diagonal elements are nearly degenerate relative to their off-diagonal elements and and this $2\times 2$ block is maximally mixed giving an approximate spectrum
\begin{align}
	M_{\chi^0_{1}} &\approx M_1 - \frac{s_\beta\pm c_\beta}{\sqrt{2}}s_W M_Z + \mathcal O\left(\frac{M_Z^2}{M_1}\right)
	\\
	M_{\chi^0_{2}} &\approx M_1 + \mathcal O\left(\frac{M_Z^2}{M_1}\right)
	\\
	M_{\chi^0_{3}} &\approx M_1 + \frac{s_\beta\pm c_\beta}{\sqrt{2}}s_W M_Z + \mathcal O\left(\frac{M_Z^2}{M_1}\right).
	\label{eq:well:temper:neutralino:mass:matrix}
\end{align}
This gives an LSP which is $\tilde B$--$\tilde h_+$ ($\tilde B$--$\tilde h_-$) for the case $\mu <0$ ($\mu >0$).

For the case $\mu <0$, observe that that as $\tan \beta \to 1$, (\ref{eq:well:tempered:compression}) fails and the diagonal elements are split by more than the off-diagonal elements. In this case the mixing remains small and the mass eigenstates are dominantly those in (\ref{eq:well:temper:b:hm:hp}). This may cause concern that the well-tempering is no longer effective since the LSP is no longer well-mixed between the `too small annihilation rate' bino and the `too large annihilation rate' Higgsinos. This is not the case, since even in the well-mixed case, the correct LSP relic abundance is obtained through coannihilation with the other neutralinos and charginos which are very close in mass to the LSP~\cite{Griest:1990kh}. In other words, the well-tempering for the LSP relic abundance and compressed spectra go hand-in-hand.

\subsection{Blind Spotting}

Thus far $\tan\beta$ has been a free parameter that controls the relative mixing of the bino with the different Higgsinos in (\ref{eq:well:temper:neutralino:mass:matrix}). It was pointed out in~\cite{Cheung:2012qy} that $\tan\beta$ controls slices of parameter space where the direct detection experiments are blind to the LSP. 
The LSP--Higgs coupling is uniquely responsible for spin-independent direct detection since Majorana particles like the neutralinos have no vector current. 
We consider a slice of parameter space where this LSP--Higgs coupling vanishes, focusing specifically on the case where 
\begin{align}
	M_1 + \mu \sin2\beta = 0.
	\label{eq:blind:spot}
\end{align}
The vanishing of the LSP coupling to the Higgs is simple to see heuristically: in this case $\mu <0$ so that the LSP is a $\tilde B$--$\tilde h_+$ mixture, following the analysis above. The relevant coupling to the Higgs comes from the $\tilde B\tilde h_{u,d} h_{u,d}$ gauge interactions. Thus the LSP--Higgs coupling comes from $\tilde B (\tilde h_u + \tilde h_d) h$ and is diagrammatically
\begin{align}
% t-chan
\begin{tikzpicture}[line width=1.35, baseline=(current  bounding  box.center), line cap=round, scale=.65]
	\tikzstyle{conjugate}=[violet!50!white]
	\tikzstyle{hard}=[line width=2, blue!60!black]
	\tikzstyle{soft}=[line width=1, dashed, red!40!white]
	\tikzstyle{tpdf}=[line width=2, dashed, green!60!black]
	\def\boxlen{1};
	\def\sholen{.75};
	\def\lablen{.8};
	\coordinate (c) at (0,0);
	\coordinate (l) at (-\boxlen,0);
	\coordinate (r) at (\boxlen,0);
	\coordinate (b) at (0,-\sholen);
	\draw[vector, line width=.75] (l) -- (c);
	\draw[line width=.75] (l) -- (c);
	\draw[vector, line width=.75] (r) -- (c);
	\draw[line width=.75] (r) -- (c);
	\draw[dashed] (c) -- (b);
	\node at (-\lablen,.5) {$ \chi_1^0$};
	\node at (\lablen,.5) {$ \chi_1^0$};
	\node at (-.3,-.6) {$h$};
\end{tikzpicture}  
&\quad\sim\quad
\begin{tikzpicture}[line width=1.35, baseline=(current  bounding  box.center), line cap=round, scale=.65]
	\tikzstyle{conjugate}=[violet!50!white]
	\tikzstyle{hard}=[line width=2, blue!60!black]
	\tikzstyle{soft}=[line width=1, dashed, red!40!white]
	\tikzstyle{tpdf}=[line width=2, dashed, green!60!black]
	\def\boxlen{1};
	\def\sholen{.75};
	\def\lablen{.9};
	\coordinate (c) at (0,0);
	\coordinate (l) at (-\boxlen,0);
	\coordinate (r) at (\boxlen,0);
	\coordinate (b) at (0,-\sholen);
	\draw[vector, line width=.75] (l) -- (c);
	\draw[line width=.75] (l) -- (c);
	\draw[] (r) -- (c);
	\draw[dashed] (c) -- (b);
	\node at (-\lablen,.5) {$\tilde B$};
	\node at (\lablen,.5) {$\tilde h_u$};
	\node at (-.3,-.6) {$h$};
	\node at (0,.4) {\textcolor{purple}{$c_{u}$}};
\end{tikzpicture} 
+
\begin{tikzpicture}[line width=1.35, baseline=(current  bounding  box.center), line cap=round, scale=.65]
	\tikzstyle{conjugate}=[violet!50!white]
	\tikzstyle{hard}=[line width=2, blue!60!black]
	\tikzstyle{soft}=[line width=1, dashed, red!40!white]
	\tikzstyle{tpdf}=[line width=2, dashed, green!60!black]
	\def\boxlen{1};
	\def\sholen{.75};
	\def\lablen{.9};
	\coordinate (c) at (0,0);
	\coordinate (l) at (-\boxlen,0);
	\coordinate (r) at (\boxlen,0);
	\coordinate (b) at (0,-\sholen);
	\draw[vector, line width=.75] (l) -- (c);
	\draw[line width=.75] (l) -- (c);
	\draw[] (r) -- (c);
	\draw[dashed] (c) -- (b);
	\node at (-\lablen,.5) {$\tilde B$};
	\node at (\lablen,.5) {$\tilde h_d$};
	\node at (-.3,-.6) {$h$};
	\node at (0,.4) {\textcolor{purple}{$c_{d}$}};
\end{tikzpicture} ,
\label{eq:top:pdf:from:gluon:splitting}
\end{align}
where $c_u$ and $c_d$ are the bino--Higgsino--Higgs gauge couplings for the up- and down-type neutral Higgsinos respectively. Recalling that the up-type and down-type Higgs supermultiplets have opposite hypercharge, $c_u = - c_d$ so that the LSP--Higgs coupling vanishes. Taking the full mixing into account, the position of the blind spot is controlled by $\tan \beta$ through the $\sin 2\beta$ in (\ref{eq:blind:spot}).

In the case where $\tan\beta \to 1$, $\sin 2\beta \to 1$ and (\ref{eq:blind:spot}) is compatible with to the well-tempered condition $M_1 \approx |\mu|$. Indeed, this also aligns with a blind spot for spin-dependent direct detection since in this limit, left-right parity is restored between the $\tilde h_{u,d}$ so that the parity-violating spin-dependent cross section vanishes. 

%\flip{Say something about `blind spot' from purity}t

\subsection{Sketched Phenomenology}

We thus consider the scenario where $\mu \to -M_1$ and $\tan\beta \to 1$ in which we live in a double-blind spot against spin-independent and spin-dependent direct detection experiments. As remarked above, there is a qualitative difference depending on which condition is more closely met: either the LSP is very well mixed between $\tilde B$ and $\tilde h_+$ or it is a pure $\tilde h_+$ state\footnote{In this latter case, the $\chi^0\chi^0 h$ coupling is still suppressed by purity~\cite{Cheung:2012qy} since both diagrams on the right-hand side of (\ref{eq:top:pdf:from:gluon:splitting}) vanish in this limit.}, depending on which of $M_1$ or $|\mu|$ is smaller. As noted, this has a small effect on the relic abundance since this is well-tempered through coannihilations with the nearly-degenerate states. This also affects the specific couplings of the electroweakinos at colliders, though we do not expect this to affect the searches we present here.  The smallness of $\tan \beta$ (combined with the decoupling of scalar superpartners) also avoids the most stringent bounds flavor-changing neutral currents such as $B_s\to\mu\mu$~\cite{Gogoladze:2010ch}.

We remark that it there is conventional wisdom that LEP excludes the MSSM with $\tan\beta \lesssim \mathcal O(\text{few})$ so that one may question the $\tan\beta \to 1$ limit taken here. This statement, however, depends on the conventional assumption that the scale of the scalar superpartners of the Standard Model fermions, $M_S$, is close to the TeV scale. This assumption is tied to the naturalness of the MSSM. In the absence of new physics, it is perhaps useful to relax any bias about the form of naturalness and instead search more broadly. This approach has recently been advocated in~\cite{Djouadi:2015jea,Djouadi:2013uqa,Djouadi:2013vqa} where it is pointed out that the low-$\tan\beta$ regime re-opens for decoupled scalar superparters---the limit we take here. The $\tan\beta\sim$ few limit is qualitatively similar and relaxes the lower limit on $M_S$.

We do not perform a detailed phenomenological study of this slice of the MSSM, such as checking precise values of the relic abundance\footnote{In non-minimal contexts, one the LSP relic abundance needn't match the observe dark matter abundance, for example if the LSP is only one component of dark matter or if one invokes non-thermal production~\cite{Cheung:2012qy}.} or bounds from indirect detection, and leave this to future work. See~\cite{Cheung:2012qy} for a detailed study and parameter scan which validates the existence of phenomenologically viable models along this region. We focus on the specific slice $\mu=- M_1$ and $\tan \beta =1$, noting that theories that are nearby behave qualitatively the same. We also present results for $\tan\beta = 15$ as a benchmark for how the model behaves for more moderate $\tan\beta$ values. See, for example~\cite{Masiero:2004ft, Bernal:2007uv} for further phenomenological considerations as one deviates from the doubly-blind spot, and~\cite{Cahill-Rowley:2014twa} for a more general status report in the phenomenological MSSM.

Finally, we note we treat the MSSM as a framework for models that populate this bino--Higgsino regime that are themselves UV complete but also span a range of phenomenology that is applicable to a broad range of models beyond the MSSM. An alternative approach is to consider minimal simplified models that can populate the analogous phenomenology, such as singlet--doublet models~\cite{Cohen:2011ec}.
In the remainder of this document, we explore the ability of collider searches to expose this difficult-to-probe region of the bino--Higgsino parameter space despite its compressed spectra.

%
%\subsection{Remarks on naturalness and generality}
%
%The bino--Higgsino model we examine here is chosen as a phenomenological model that can probe a hard-to-test region of parameter space. We have completely neglected the ingredients of the MSSM which address the Hierarchy problem and have assumed these to decouple. This regime can be realized in high-scale SUSY models such as split-SUSY~\cite{Giudice:2004tc,ArkaniHamed:2004fb}

\section{Signal generation}

The parameters \mone, \mtwo, $\mu$ and $\tan\beta$ control the mixings, couplings, and masses 
of the gauginos. These parameters are generated along a line 
\begin{align}
	M_1 = -\mu,
	\label{eq:well:tempered:Higgsino:Bino}
\end{align}
%$-M_1=\mu$, 
with $\tan\beta=$ 1 and 15 and decoupled wino,
$M_2=4000~\text{GeV} \gg M_1$. This gives a well-tempered bino-Higgsino LSP.
Multi-jet final states produced in association with two gauginos are generated at $\sqrt{s}=8$ TeV. Samples are generated with zero, one, and two additional hard partons using MadGraph5~\cite{madgraph} and the underlying event and particle showering is modeled with Pythia~8~\cite{pythia}. The VBF and Drell-Yan components are generated simultaneously. Drell-Yan dominates the production cross section.

The kinematics of the jets in VBF events depend on the mass of the sparticles produced. As heavier  sparticles are produced by vector boson fusion, more of the energy from the bosons must go into the  sparticle mass, providing them with less momenta. Therefore, the $\Delta R$ between the two jets in the  final state increases as the mass of the particles in the final state increases. 

\begin{figure*}[htp]
\begin{center}
\subfigure[$\tan\beta$=1.]{
\includegraphics[scale=0.38]{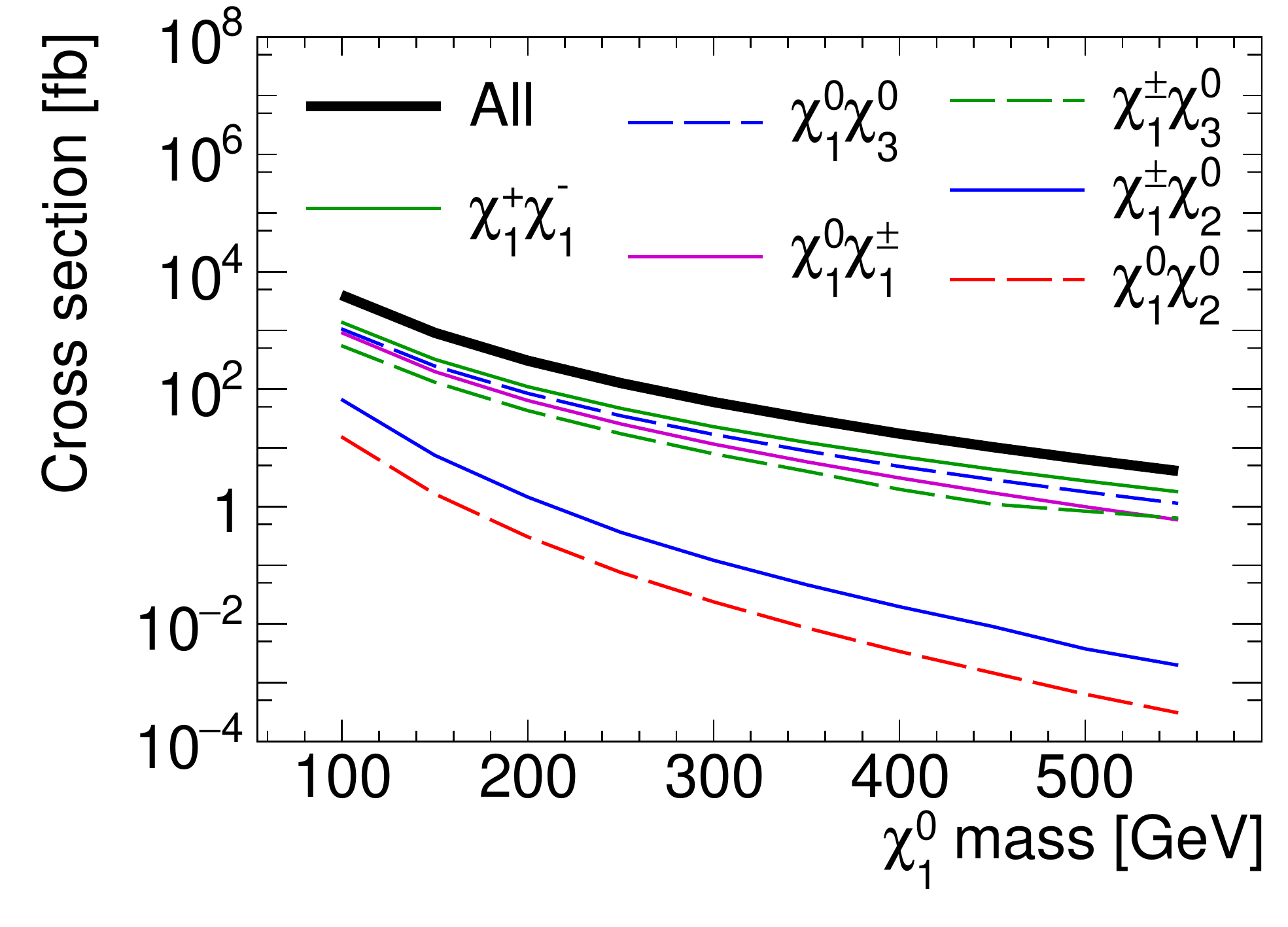}
}\quad
\subfigure[$\tan\beta$=15.]{
\includegraphics[scale=0.38]{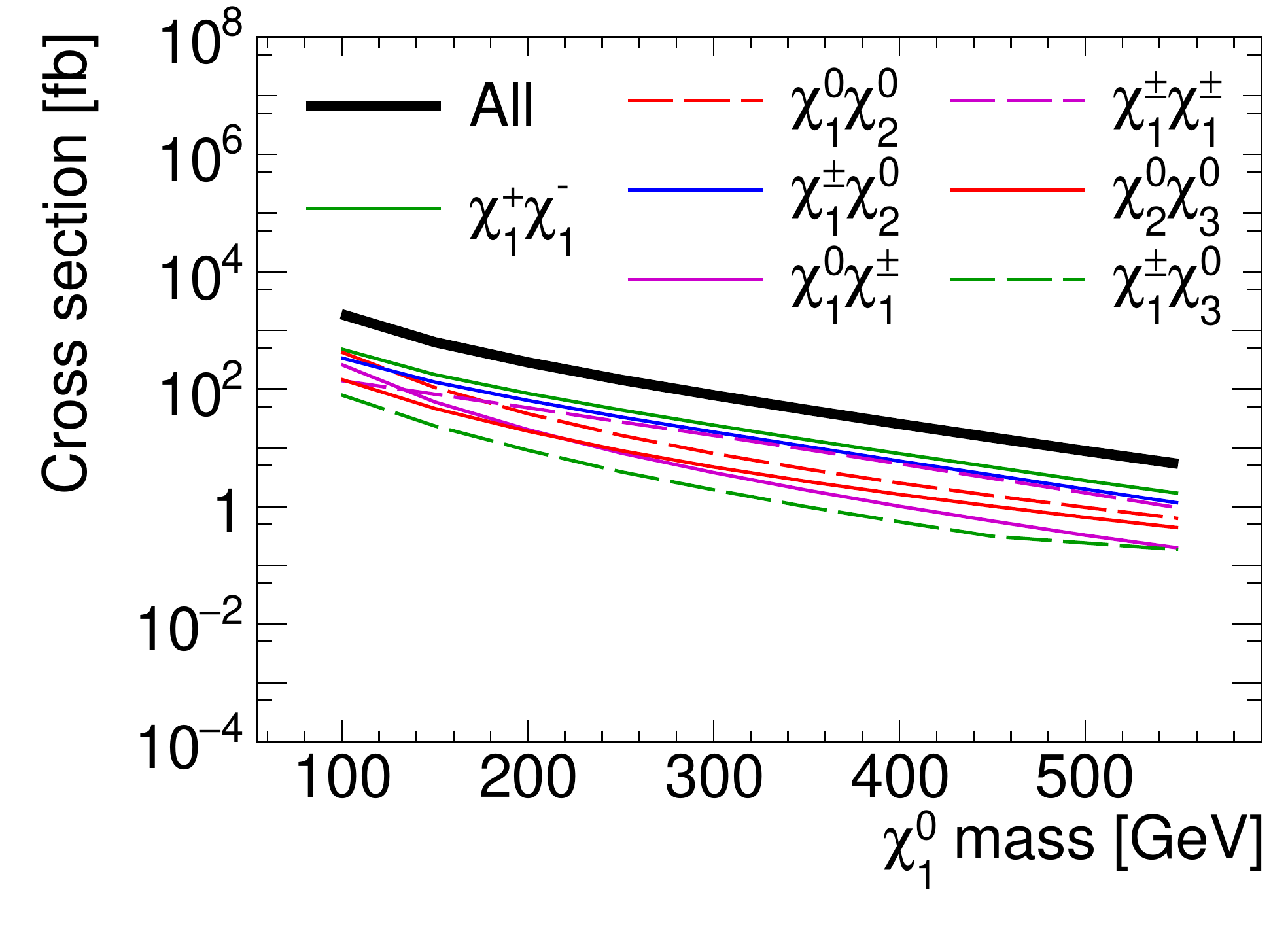}
}
\caption{\label{fig:prodXsec}Production cross sections in $pp$ collisions at $\sqrt{s}=8$ TeV at leading order for models of compressed spectra as a function of the mass of the lightest SUSY particle, $\chi_1^0$, for various combinations of gaugino modes. Other gaugino modes in which the cross sections are significantly smaller are not considered.}
\end{center}
\end{figure*}

The production cross sections are shown in Figure~\ref{fig:prodXsec}.

\section{Analysis}

Two ATLAS datasets are reinterpreted to constrain the models described above. The first is a broad ATLAS search for new phenomena in events with one or more jet and large missing transverse momentum (``mono-jet'')~\cite{atlasMonojet}. The second is a search for vector-boson fusion (VBF) production of a Higgs boson which subsequenetly decays invisibly, giving two forward jets and large missing transverse momentum~\cite{atlasVBF}.

\subsection{Mono-jet Selection}

The mono-jet search~\cite{atlasMonojet} defines accepted jets as those with $p_{\rm T}>30$ GeV 
and $|\eta|<4.5$.
The leading jet must satisfy $p_{\rm T}>120$ GeV and $|\eta|<2$.
The ratio of the leading jet transverse momentum and the missing transverse momentum must be 
greater than 0.5. 
No jet may be within $\Delta\phi=1$ of the missing transverse momentum.
There are nine signal regions characterized by their increasingly tighter requirements on the 
missing transverse momentum, $\missET>$ 150, 200, 250, 300, 350, 400, 500, 600,
700 GeV. 
Events with leptons are vetoed.

\begin{figure*}[htp]
\begin{center}
\subfigure[$\tan\beta$=1.]{
\includegraphics[scale=0.38]{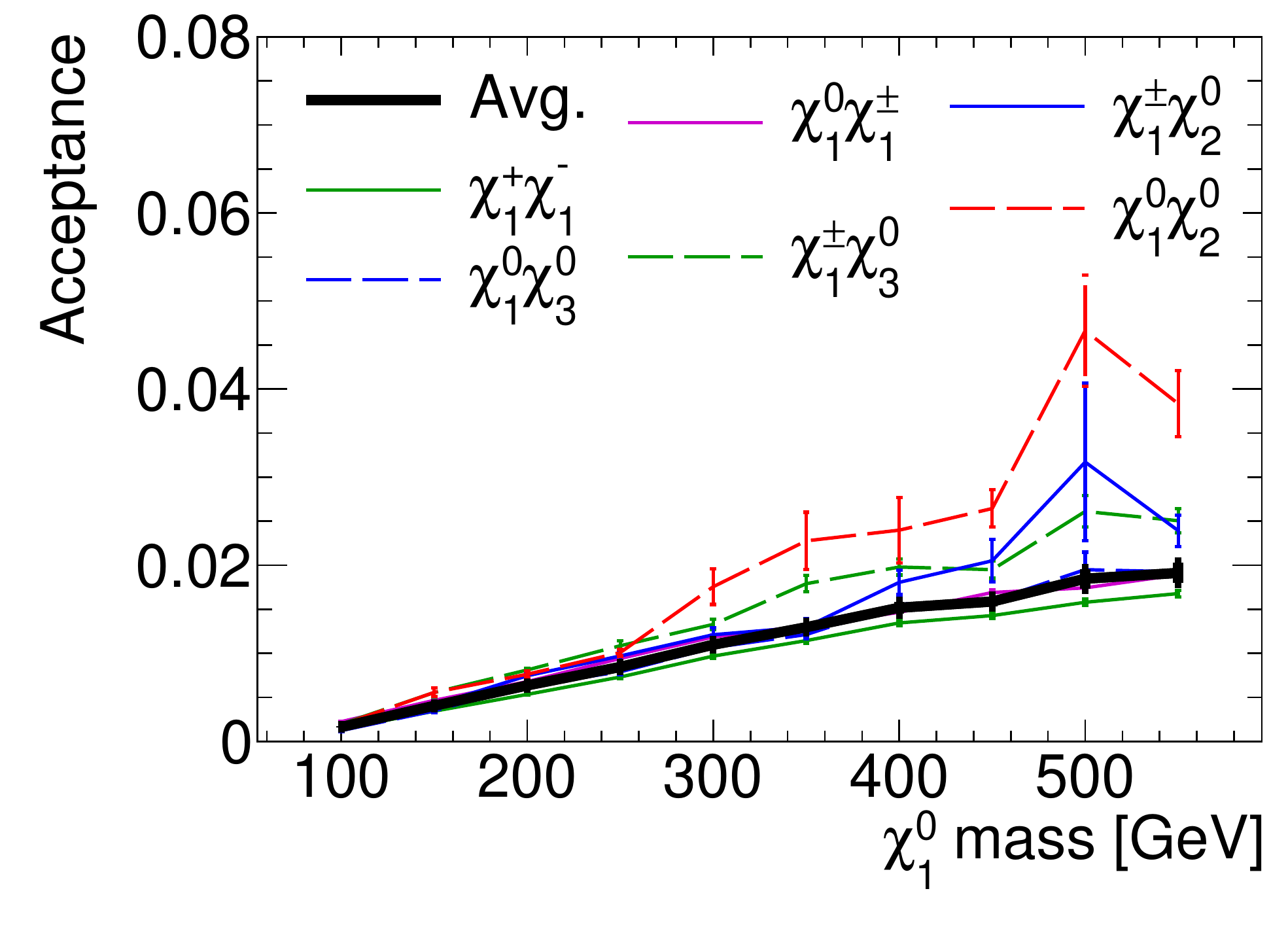}
}\quad
\subfigure[$\tan\beta$=15.]{
\includegraphics[scale=0.38]{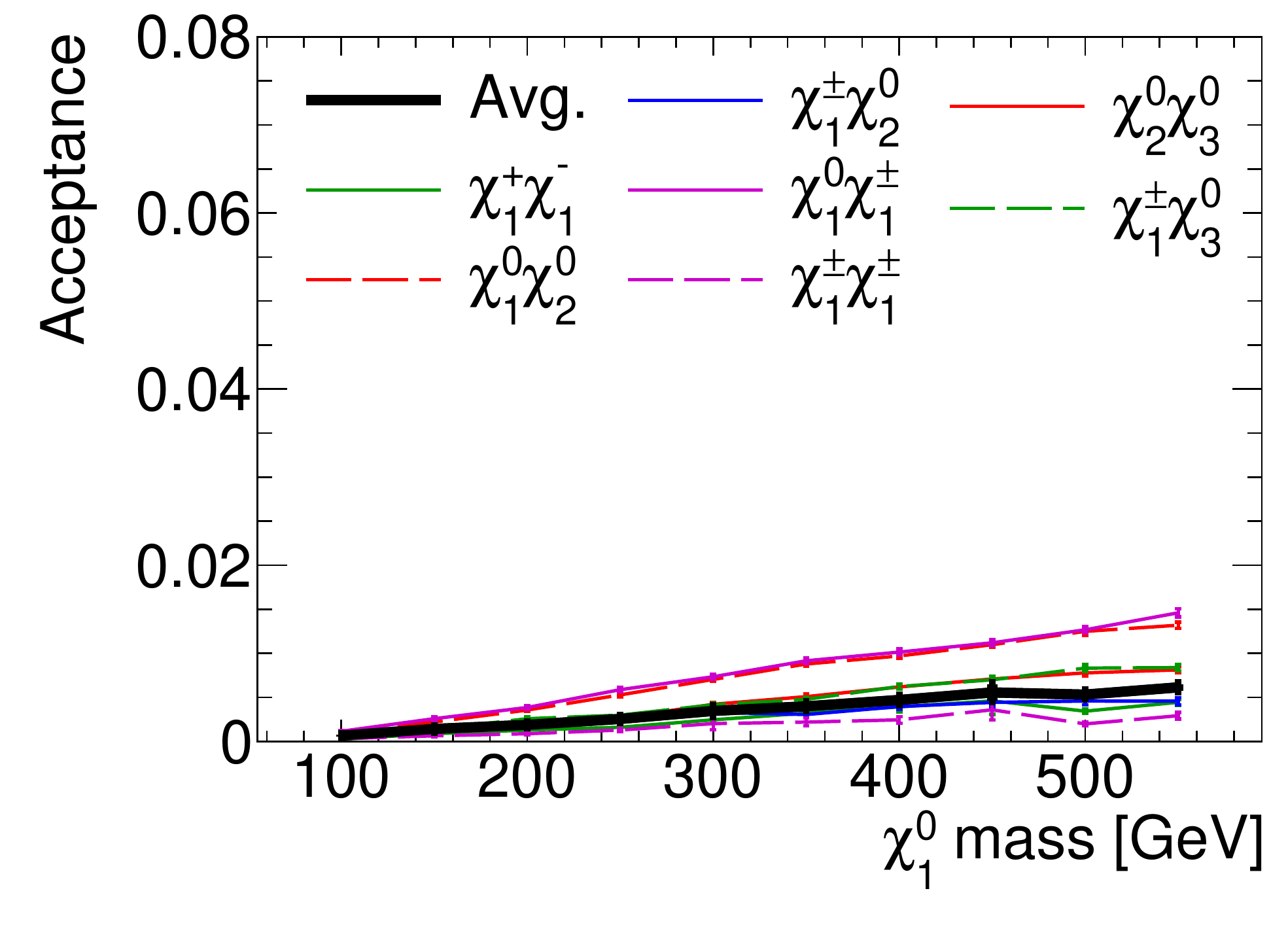}
}
\caption{\label{fig:monojetAcc}Acceptance of the mono-jet selection for $\missET>$ 400 GeV is measured with simulated events  at $\sqrt{s}=8$ TeV at leading order for models of compressed spectra as a function of the mass of the lightest SUSY particle, $\chi_1^0$, for various combinations of gaugino modes. Also shown is the cross-section weighted average. }
\end{center}
\end{figure*}

The acceptance, the fraction of events which survive this selection, is shown for each production case in Figure~\ref{fig:monojetAcc}. The 
cross-section-weighted average is also shown.

\subsection{VBF Selection}

The VBF analysis~\cite{atlasVBF} requires missing transverse momentum be greater than 150 GeV, 
the leading 
and sub-leading jet $p_{\rm T}$ be greater than 75 and 50 GeV, respectively. 
The product of the leading and sub-leading jets' $\eta$ must be less than zero. The 
absolute value of the difference in the leading and sub-leading jets' $\eta$ and $\phi$ must 
be greater than 4.8 and less than 2.5, respectively. The mass of the jet pair must be greater 
than 1 TeV. 
No jet may be within $\Delta\phi=1$ of the missing transverse momentum.
Any events with more than 2 jets with $p_{\rm T}>30$ GeV and $|\eta|<4.5$ are vetoed.
Events with leptons are vetoed. 

\begin{figure*}[htp]
\begin{center}
\subfigure[$\tan\beta$=1.]{
\includegraphics[scale=0.38]{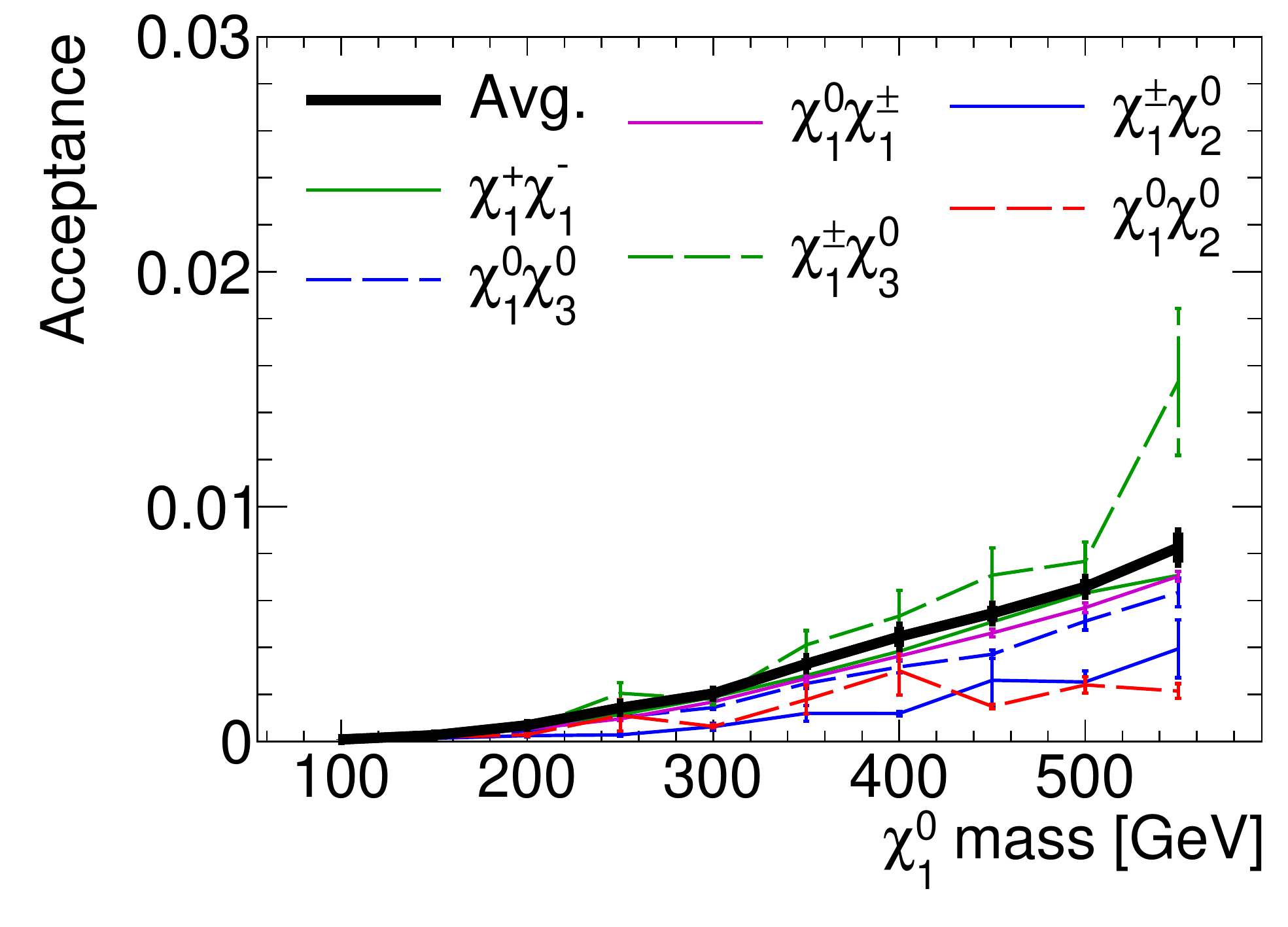}
}\quad
\subfigure[$\tan\beta$=15.]{
\includegraphics[scale=0.38]{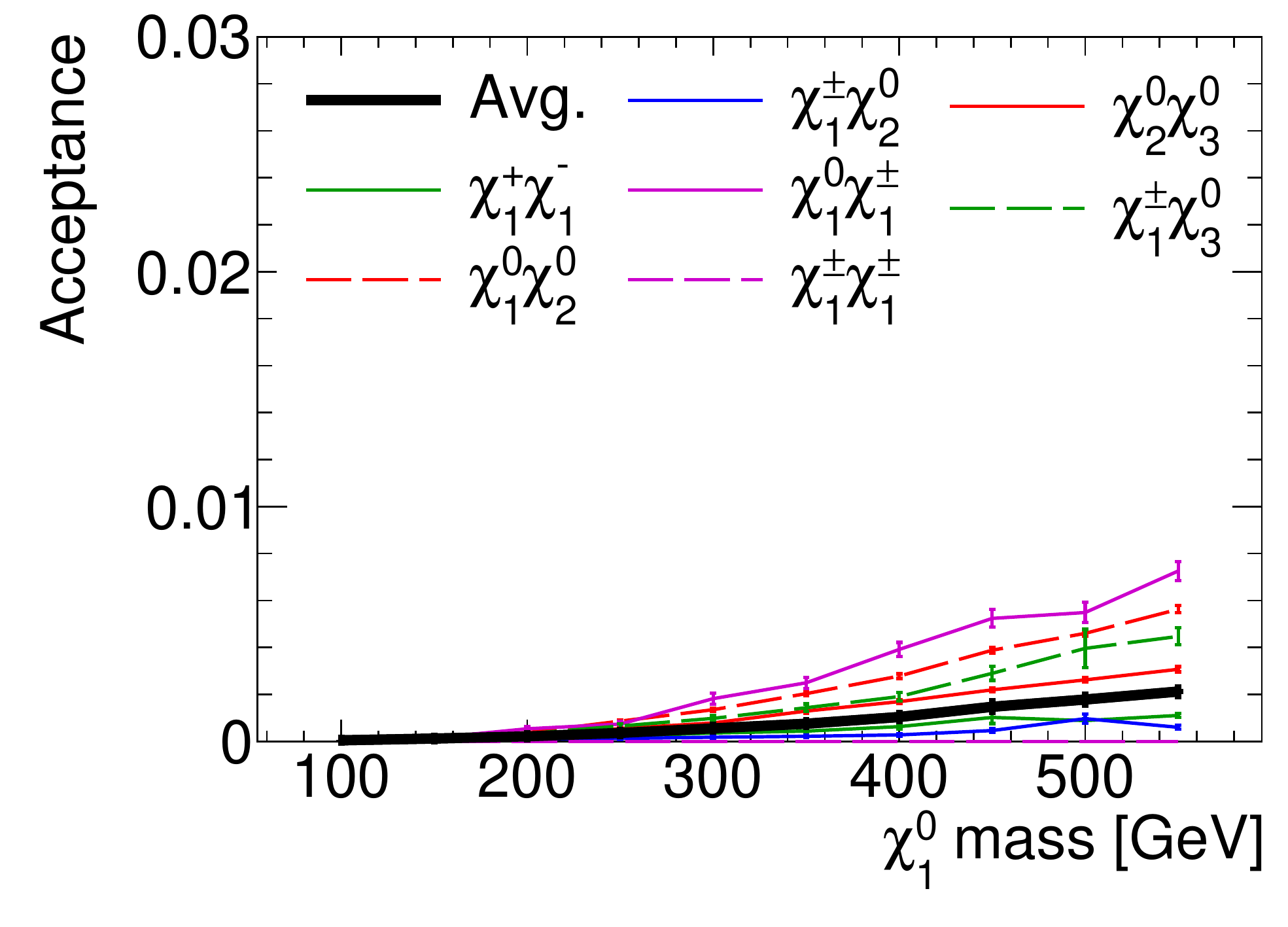}
}
\caption{\label{fig:dijetAcc}Acceptance of the VBF selection measured with simulated events  at $\sqrt{s}=8$ TeV at leading order for models of compressed spectra as a function of the mass of the lightest SUSY particle, $\chi_1^0$, for various combinations of gaugino modes. Also shown is the cross-section weighted average.}
\end{center}
\end{figure*}

The acceptance, the fraction of events which survive this selection, is shown for each production case in Figure~\ref{fig:dijetAcc}. The cross-section-weighted average is also shown.

\section{Model reinterpretation}

\subsection{Procedure}

In both the mono-jet and VBF cases, ATLAS provides model-independent upper limits on the visible production cross section:

\[ \sigma_{\textrm{vis}} = \sigma \times \epsilon \times \cal{A} \]

\noindent
where the acceptance $\cal{A}$ is the fraction of generated events which fall into the specified fiducial region and $\epsilon$ is the reconstruction efficiency inside the fiducial region.  Typically the model dependence is captured by $\cal{A}$ and can be measured at the parton-level, while $\epsilon$ is fairly model-independent and reported by the experiment.   Applying the dataset to our models therefore only requires measuring $\cal{A}_{\rm SUSY}$ so that one can calculate:

\[\sigma^{\rm limit}_{\rm SUSY} = \frac{\sigma_{\rm vis}}{\cal{A}_{\rm SUSY}\times\epsilon}. \]

\subsection{Mono-jet Dataset}

The mono-jet analysis reports limits for nine different signal regions defined by different $E^{\rm miss}_{\rm T}$ requirements; see Tab.~\ref{tab:monojetLimits}. The reconstruction efficiency $\epsilon$ is reported for four different
signal regions and ranges from 81\% to 88\%. We use a linear interpolation to calculate the efficiencies for the  intermediate regions.  We choose the signal region which gives the tightest expected limits in each case. See Tab.~\ref{tab:res} for details.

\begin{table}
\begin{center}
\caption{\label{tab:monojetLimits} Observed and expected upper limits on $\sigma_{\textrm{vis}}$ at 95\% CL from the ATLAS mono-jet dataset~\cite{atlasMonojet}.} 
\begin{tabular}{l|r|r}
\hline
\hline
& Observed & Expected \\
$E^{\rm miss}_{\rm T}>$ &  limit [fb] &  limit [fb] \\
\hline
$150$ GeV & 726 & 935 \\
\hline
$200$ GeV & 194 & 271 \\
\hline
$250$ GeV &  90 & 106 \\
\hline
$300$ GeV &  45 &  51 \\
\hline
$350$ GeV &  21 &  29 \\
\hline
$400$ GeV &  12 &  17 \\
\hline
$500$ GeV & 7.2 & 7.2 \\
\hline
$600$ GeV & 3.8 & 3.6 \\
\hline
$700$ GeV & 3.4 & 1.8 \\
\hline
\hline
\end{tabular}
\end{center}
\end{table}

\subsection{VBF Dataset}

An expected (observed) upper limit  on $\sigma_{\textrm{vis}}$ at 95\% confidence level is reported as 4.8 (3.9) fb from the VBF dataset. The reconstruction efficiency $\epsilon$ is 94\%~\cite{atlasVBF}. Together with $\cal{A}_{\textrm{SUSY}}$ this allows calculation of upper limits on the SUSY models described above. 
  
\subsection{Results}

\begin{figure*}[htp]
\begin{center}
\subfigure[$\tan\beta$=1.]{
\includegraphics[scale=0.38]{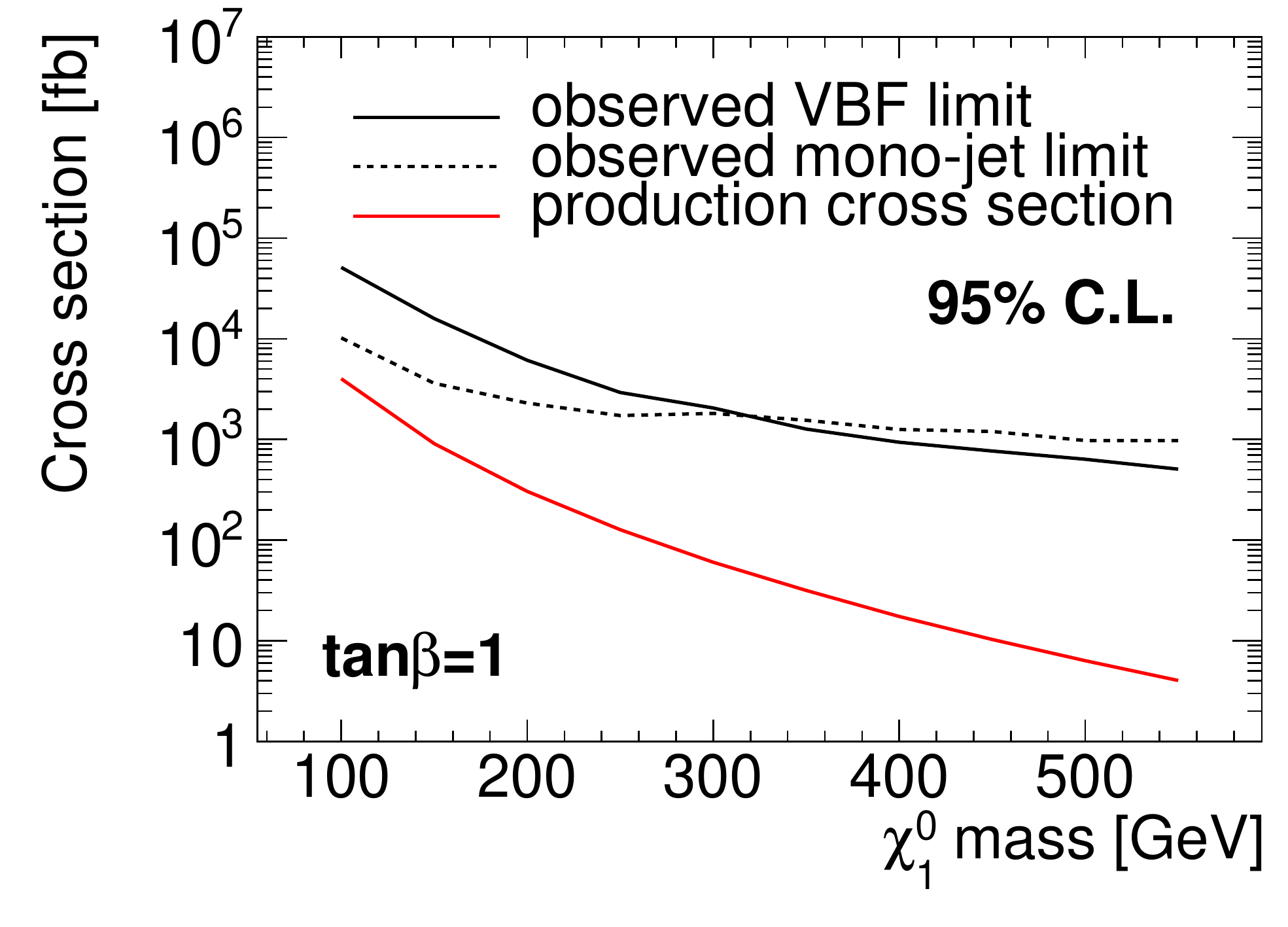}
}\quad
\subfigure[$\tan\beta$=15.]{
\includegraphics[scale=0.38]{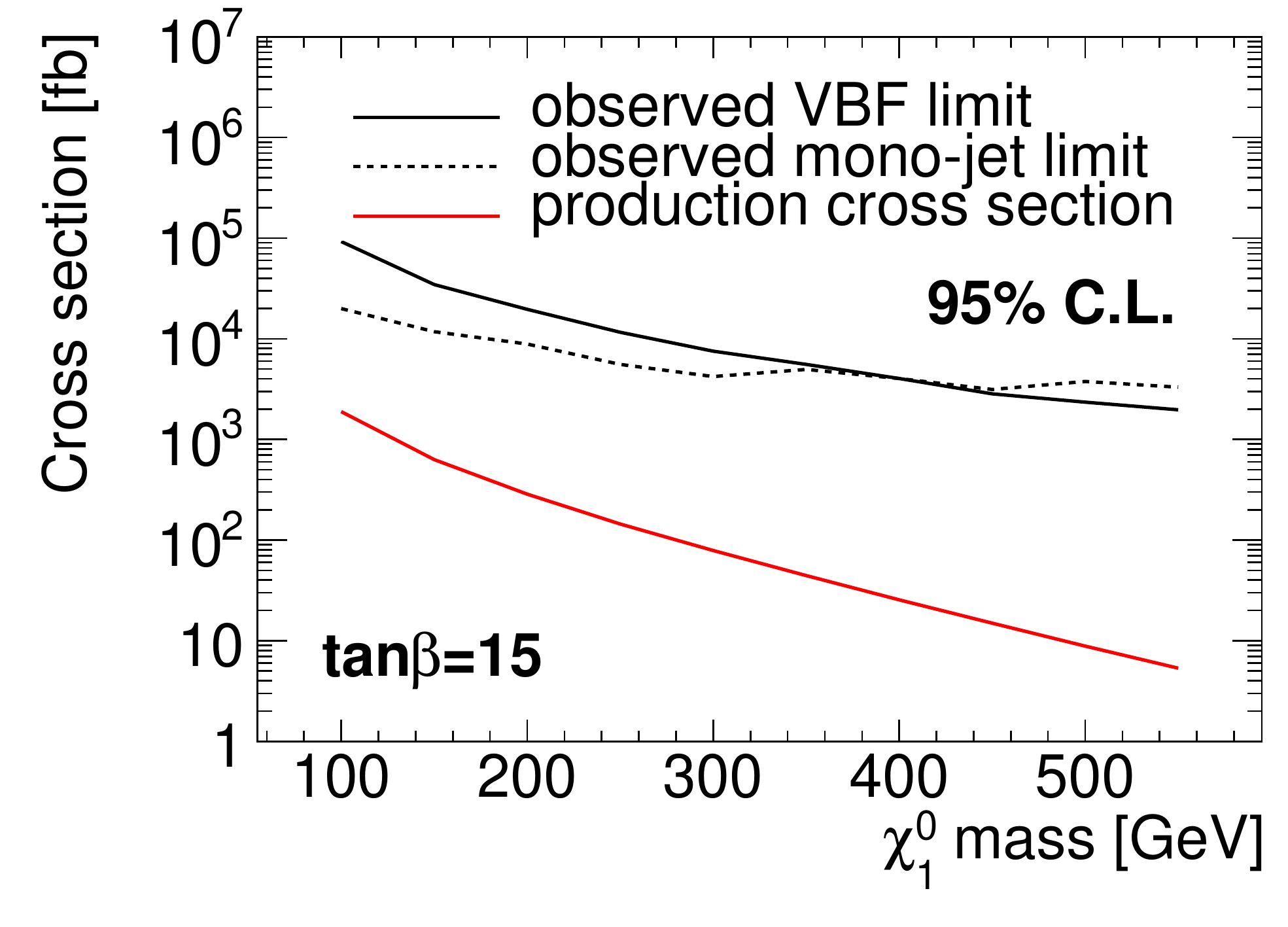}
}
\caption{\label{fig:combinedLimits} Limits at 95\% confidence level on the production cross section of compressed spectra models as a function of the mass of the lightest SUSY particle using the ATLAS VBF and mono-jet datasets.}
\end{center}
\end{figure*}

Upper limits on the pair-production of gauginos in association with jets from the mono-jet and VBF datasets are shown in Fig~\ref{fig:combinedLimits} and Table~\ref{tab:res} along with the theoretical production cross-section at leading order.

\begin{table*}
\caption{ For $\tan\beta=$1, details of the theory production cross section $\sigma_{\textrm{Theory}}$, and for both the mono-jet and VBF datasets: the acceptance $\cal{A}$, reconstruction efficiency $\epsilon$, and 95\% CL upper limit $\sigma^{\textrm{Limit}}$ for various choices of the $\chi_1^0$ mass. In the mono-jet case, the selected signal region (see Tab.~\ref{tab:monojetLimits}) is also indicated.}
\label{tab:res}
\begin{tabular}{l|r|rrrr|rrr}
\hline\hline
$m_{\chi^0_1}$& $\sigma_{\textrm{Theory}}$ & \multicolumn{4}{c|}{mono-jet} & \multicolumn{3}{c}{VBF}\\
 (GeV) &  (fb) & SR$_{\textrm{mono-jet}}$ & $\cal{A}_{\textrm{mono-jet}}$ & $\epsilon_{\textrm{mono-jet}}$ & $\sigma^{\textrm{Limit}}_{\textrm{mono-jet}}$ (fb) & $\cal{A}_{\textrm{VBF}}$ & $\epsilon_{\textrm{VBF}}$ & $\sigma^{\textrm{Limit}}_{\textrm{VBF}}$ (fb) \\
\hline
100 & 4010 & 4 & 0.0053 & 0.83 & 10000 & $8.2\times10^{-5}$ & 0.94 & 51000 \\
150 &  908 & 6 & 0.0040 & 0.83 &  3600 & $2.6\times10^{-4}$ & 0.94 & 16000 \\
200 &  305 & 6 & 0.0064 & 0.83 &  2300 & $6.8\times10^{-4}$ & 0.94 &  6200 \\
250 &  127 & 6 & 0.0085 & 0.83 &  1700 & $1.4\times10^{-3}$ & 0.94 &  2900 \\
300 & 60.2 & 7 & 0.0048 & 0.82 &  1800 & $2.1\times10^{-3}$ & 0.94 &  2000 \\
350 & 31.6 & 7 & 0.0056 & 0.82 &  1600 & $3.3\times10^{-3}$ & 0.94 &  1300 \\
400 & 17.4 & 7 & 0.0069 & 0.82 &  1300 & $4.3\times10^{-3}$ & 0.94 &   970 \\
450 & 10.3 & 7 & 0.0073 & 0.82 &  1200 & $5.4\times10^{-3}$ & 0.94 &   770 \\
500 &  6.3 & 7 & 0.0089 & 0.82 &   990 & $6.7\times10^{-3}$ & 0.94 &   630 \\
550 &  4.0 & 7 & 0.0090 & 0.82 &   970 & $8.2\times10^{-3}$ & 0.94 &   510 \\
\hline
\hline
\end{tabular}
\end{table*}

\begin{table*}
\caption{ For $\tan\beta$=15, details of the theory production cross section $\sigma_{\textrm{Theory}}$, and for both the mono-jet and VBF datasets: the acceptance $\cal{A}$, reconstruction efficiency $\epsilon$, and 95\% CL upper limit $\sigma^{\textrm{Limit}}$ for various choices of the $\chi_1^0$ mass. In the mono-jet case, the selected signal region (see Tab.~\ref{tab:monojetLimits}) is also indicated.}
\label{tab:res}
\begin{tabular}{l|r|rrrr|rrr}
\hline\hline
$m_{\chi^0_1}$& $\sigma_{\textrm{Theory}}$ & \multicolumn{4}{c|}{mono-jet} & \multicolumn{3}{c}{VBF}\\
 (GeV) &  (fb) & SR$_{\textrm{mono-jet}}$ & $\cal{A}_{\textrm{mono-jet}}$ & $\epsilon_{\textrm{mono-jet}}$ & $\sigma^{\textrm{Limit}}_{\textrm{mono-jet}}$ (fb) & $\cal{A}_{\textrm{VBF}}$ & $\epsilon_{\textrm{VBF}}$ & $\sigma^{\textrm{Limit}}_{\textrm{VBF}}$ (fb) \\
\hline
100 & 1890 & 3 & 0.0054 & 0.83 & 20000 & $4.5\times10^{-5}$ & 0.94 & 93000 \\
150 &  630 & 3 & 0.0092 & 0.83 & 12000 & $1.2\times10^{-4}$ & 0.94 & 35000 \\
200 &  286 & 4 & 0.0061 & 0.83 &  8900 & $2.1\times10^{-4}$ & 0.94 & 20000 \\
250 &  144 & 5 & 0.0046 & 0.83 &  5600 & $3.6\times10^{-4}$ & 0.94 & 12000 \\
300 & 78.8 & 6 & 0.0035 & 0.82 &  4200 & $5.6\times10^{-4}$ & 0.94 &  7500 \\
350 & 44.4 & 7 & 0.0018 & 0.82 &  5000 & $7.5\times10^{-4}$ & 0.94 &  5600 \\
400 & 25.5 & 7 & 0.0022 & 0.82 &  4000 & $1.0\times10^{-3}$ & 0.94 &  4000 \\
450 & 14.9 & 7 & 0.0028 & 0.82 &  3100 & $1.5\times10^{-3}$ & 0.94 &  2800 \\
500 &  8.8 & 7 & 0.0023 & 0.82 &  3800 & $1.8\times10^{-3}$ & 0.94 &  2300 \\
550 &  5.3 & 7 & 0.0027 & 0.82 &  3300 & $2.1\times10^{-3}$ & 0.94 &  2000 \\
\hline
\hline
\end{tabular}
\end{table*}

\section{Discussion}

This is the first application of the VBF dataset to these SUSY scenarios.
Neither dataset provides limits within an order of magnitude of the theoretical prediction, highlighting the challenge of probing these SUSY scenarios. However, we note that the mono-jet and VBF datasets have complementary sensitivity, as mono-jet dominates at lower $\chi_0$ mass and the VBF dataset is more powerful for more massive $\chi_0$.  In the case of $\tan\beta=1$, the limits are within a factor of five of the theoretical prediction.  Similar searches at $\sqrt{s}=13$ TeV will yield considerably more stringent limits.

%More clever insights here. \emph{these are spit-balls, polish and include them as you see fit}
%\begin{enumerate}
%	\item Future direction: quantify the ``alive'' region outside of $M_1 = -\mu$ and $\tan \beta = 1$ where there are parameter points which still exist. For example this seems constrained in the pMSSM~\cite{Cahill-Rowley:2014twa}. This probably wouldn't be hard, we can make sure of DarkSUSY to filter parameter points. It could be done with an patient grad student. 
%	\item Future direction: do a parameter scan in a simplified model that mocks up these interactions. We could consider singlet--doublets (e.g.~\cite{Cohen:2011ec}), or more generally consider WIMPy-models with mediators that hide really well and nearly-degenerate dark matter states. 
%	\item Future direction: explore other blind spots. For example, bino--wino mixing, or the blind spot when the heavy and light Higgs contributions are tuned to cancel the spin-independent cross section. This latter case was mentioned in the blind-spot paper as something that could be explored but that they didn't do.
%\end{enumerate} 

\section*{Acknowledgements}

We thank Aaron Pierce,
Joanne Hewett,
Josh Ruderman,
and
Tim M.P.~Tait
for useful discussions.
The work of D.W.\ and P.T.\ was performed in part at the Aspen Center for
Physics, which is supported by National Science Foundation grant
PHY--1066293.
P.T.~is supported in part by a UCI Chancellor's ADVANCE fellowship.
This work is supported in part by NSF Grants No.~PHY--1316792. D.W.\ and A.N.\ are supported by a grant from the DOE Office of Science.

\bibliography{compressedSpectra}

\end{document}